\documentclass[english]{article}
\usepackage[T1]{fontenc}
\usepackage[latin9]{inputenc}
\usepackage{amsmath}
\usepackage{babel}

\begin{document}

\title{A probabilistic model to describe the dual phenomena of biochemical
pathway damage and biochemical pathway repair.}

\date{\textbf{Anirban Banerji}\\
Bioinformatics Centre, University of Pune, Pune-411007, Maharashtra,
India.\textbf{}\\
\textbf{E-mail address : anirbanab@gmail.com}}
\maketitle
\begin{abstract}
Biochemical pathways emerge from a series of Brownian collisions between
various types of biological macromolecules within separate cellular
compartments and in highly viscous cytosol. Functioning of biochemical
networks suggests that such serendipitous collisions, as a whole,
result into a perfect synchronous order. Nonetheless, owing to the
very nature of Brownian collisions, a small yet non-trivial probability
can always be associated with the events when such synchronizations
fail to emerge consistently; which account for a damage of a biochemical
pathway. The repair mechanism of the system then attempts to minimize
the damage, in the pursuit to bring restore the appropriate level
of synchronization between reactant concentrations. Present work presents
a predictive probabilistic model that describes the various facets
of this complicated and coupled process(damaging and repairing). By
describing the cytosolic reality of Brownian collisions with Chapman-Kolmogorov
equations, the model presents analytical answers to the questions,
with what probability a fragment of any pathway may suffer damage
within an arbitrary interval of time? and with what probability the
damage to a pathway can be repaired within any arbitrary interval
of time?\\
\\
\\

\end{abstract}
Biochemical pathways come to existence due to series of concentration-dependent
collisions among various species of biological macromolecules that
constitute a pathway. Traditionally, the time evolution of biochemical
pathways is desrcibed by a set of coupled (first order) differential
equations that stem from law of mass action and the information regarding
concentrations of each species. Law of mass action is an empirical
law that connects reaction rates with molecular component concentrations
through a simple equation. Once provided with the information of initial
molecular concentrations, law of mass action presents a complete picture
of the component concentrations at all future time points (Espenson,
1995). Although popular, this approach (based on law of mass action)
assumes that process of initiating and sustaining the chemical reactions
is continuous and deterministic (Cox, 1994). As one studies smaller
and smaller systems, the validity of a continuous approach becomes
ever more tenuous and it becomes clear that in reality, chemical reactions
are innately stochastic (and not continuous) in nature. One alsorealizes
that reactions occur as discrete events resulting from random (and
not deterministic) molecular collisions (Gillespie, 1977; McAdams
and Arkin, 1999; Gibson, 2000; Golightly and Wilkinson, 2006). The
stochastic approach attempts to describe this inherent random nature
of microscopic molecular collisions to construct a probabilistic model
of the reaction kinetics (Resat et al., 2001; Qian and Elson, 2002).
This approach is thus suited to the modeling of small, heterogenous
environments typical of in-vivo conditions (Kuthan, 2001). Such intrinsically
stochastic nature of biochemical reactions have profound implications
in many spheres of biology. For example, in the paradigm of molecular
binding and chemical modifications, stochastic collisions give rise
to temporal fluctuations and cell-to-cell variations in the number
of molecules of any given type; they mask genuine signals and responses
and furthermore, contribute critically to the phenotypic diversity
in a population of genetically identical individuals (Raser and O'Shea,
2004, Maamar et al. 2007).\\
\\
Biochemical pathways are structures that depend crucially upon
accurate synchronizations between concentration profiles. But stochastic
collisions, by their very nature, are probabilsitic (Calef and Deutch,
1983). Furthermore, at the molecular level, random fluctuations are
inevitable; with their effect being most significant when molecules
are at low numbers in the biochemical system (Turner et al., 2004).
Therefore, the event of a biochemical pathway malfunctioning can well
be attributed to a failure to ensure synchronization between various
macromolecular concentrations (Magarkar et al., 2011); which in turn,
can be attributed to the inherently probabilsitic nature of the stochastic
collisions. Although these (serendipitous) Brownian collisions account
for the emergence of intricate and exquisite order in biochemical
reactions in most of the cases, the probability of their failing to
achieve the same is cannot merely be trivial. Pivotally important
to stochastic modelling is the realization that molecular reactions
are essentially random processes and therefore, it is impossible to
predict with complete certainty the time instance at which the next
reaction within any volume may take place (Turner et al., 2004). In
macroscopic systems, in the presence of a large number of interacting
molecules within the confinement of a cellular compartment, the randomness
of this behaviour averages out; hence the gross macroscopic state
of the system appears to be deterministic and predictable (Minton,
1993; Ahn et al., 1999; Ellis, 2001). However, while studying the
same from bottom-up approach, one cannot resort to macroscopic determinism
observed at the limiting case (high concentration of the interacting
macromolecules, highly viscous cytoplasmic fluid, etc.)(Gillespie,
1977; Rao and Arkin, 2003). Thus, modeling biochemical reactions from
bottom-up perspective needs to take into account the probability-driven
nature of macromolecular interactions.\\
\\
The present work assumes that owing to inherently probabilistic
nature of the collisions amongst macromolecules, the adequate level
of synchronization amongst interacting species (that is required to
ensure the emergence of macroscopic deterministic profile) - will
fail at times. We hypothesize that it is due to the failure to ensure
the appropriate extent of synchronization that a (fragment or the
entire) biochemical pathway will fail to function. A (fragment or
the entire) pathway with such incorrect synchronization is referred
to as 'damaged' pathway, in the present work. Though evolution has
given rise to robustness of biochemical pathways, it is difficult
to assume that any arbitrarily chosen pathway will always be functioning
with exactly the expected level of optimality. - Though this seems
intuitive to appreciate, one will fail to find either a theoretical
or an empirical answer to any of the two questions; one, how many
times, within a given interval of time, a pathway (or any section
of it) suffers from damage? Two, how soon will the damaged section
of the damaged pathway be repaired? etc.. An easy approach to these
simple questions may suggest that the answer to the aforementioned
question will be, first: pathway-specific, two: time-interval specific,
three: organism-specific. However, since evolution tends to reuse
the tried-and-tested mechanisms, there is reason to expect that the
answers to the aforementioned questions may not be case-specific but
general. Therefore, from a ageneral perspective, the present model
attempts to quantify one: the probability with which a pathway (or
a fragment of it) will malfunction within any arbitrarily chosen time
interval, suitable to observe such event; and two: the probability
with which the damaged pathway (or the damaged fragment of it) will
be repaired within any arbitrarily chosen time interval, suitable
to observe such event. Though attempts of probabilistic modeling of
biochemical systems are not entirely commonplace, some previous attempts
in the similar lines can be found in (Hume, 2000; Elowitz et al.,
2002; Golightly and Wilkinson, 2006).\\
\\
The present work studies two cases; first-case, when the damaged
fragment of \textbf{P} is detected immediately and repairing of this
fragment of \textbf{P} starts without any delay, second-case, when
the damaged fragment of \textbf{P} is detected after a certain time
lag and repairing of this fragment of \textbf{P}, accordingly, starts
with a delay. Though no concrete piece of data either supports or
contradicts the first case; the facts that, one: underlying mechanism
of the pathway functioning is rooted in Brownian collisions and therefore
is often unreliable, and two: even though a pathway functions due
to series of favorable but essentially serendipitous collisions, we
do not suffer from too many instances of pathway damage - suggests
that probably, after the damage of a fragment of a pathway (when concentrations
of consecutive species of macromolecules (that constitute this fragment)
fail to ensure the appropriate coupling strength, whereby at least
one reaction fails to occur optimally), the detection and subsequent
repairment of that fragment (restoring back the adequate coupling
strength between concentrations of consecutive species of macromolecules)
- take place without any appreciable passage of time. The second case,
of course, does not discuss such idealistic scenario; instead, it
attempts to model the case when damage to a part of a pathway is detected
after an appreciable time-lag, whereby the repairing mechanism starts
to work only after an appreciable passage of time.\\
\\
Damage to a biochemical pathway though possible(as argued beforehand)
are assumed to be not entirely common. Assuming that only rarely and
accidentally does the synchronization among concentrations of interacting
macromolecular species fail to satisfy the required optimality (and
therefore cause the damage to the pathway), occurrences of such sub-optimal
synchronization in any part of a biochemical pathway (\textbf{P})
are assumed to take place as a Poisson process, characterized by an
elementary flow with intensity $\lambda$. For the first case, the
lack of synchronization in any part of \textbf{P} is detected immediately
and the process of repairement of it starts without a delay. Using
similar logic as the aforementioned one, we assume that the time required
to repair the damaged sub-pathway can be described with a distribution
of exponential nature with a parameter $\mu$, whence the recovery
process can be described as :\\
\begin{equation}
f(t)=\mu e^{-\mu t}(t>0).\end{equation}
\\
\textbf{\underbar{Case -1):}} \\
For the first case, the repairing process starts as soon as the
detection of the damage takes place, which in turn, is assumed to
take place immediately as the damage takes place. Hence, the variable
t (viz. time) in eqn-1 begins from the point of detection of damage,
which implies that receovery process is described strictly in time
range $(t>0)$. Before this, viz. at the initial moment $(t=0)$,
the biochemical pathway (\textbf{P}) is assumed to be functioning
without problem. We attempt to find at first, the probability that
at any arbitrarily chosen instance t, the pathway \textbf{P} is functioning
properly and then, the probability that during any arbitrarily chosen
time interval $(0,t)$, \textbf{P} falters from its optimal functionaing
at least once; before attempting to evaluate the limiting probabilities
of the states of \textbf{P}.\\
\\
We denote the states of \textbf{P} as, $(s_{0})$, when it is
functioning properly and $(s_{1})$, when at least one part of it
is malfunctioning and is being repaired; correspondingly, the probabilities
$p_{0}$ and $p_{1}$ are assigned respectively. \\
\\
The Chapman-Kolmogorov equations (Sigman, hypertext link; Weisstein,
hypertext link) for these states, viz. $\left(p_{0}\left(t\right)\right)$
and $\left(p_{1}\left(t\right)\right)$ can be constructed as :\\
\begin{equation}
\frac{dp_{0}}{dt}=\mu p_{1}-\lambda p_{0}\end{equation}
and\\
\begin{equation}
\frac{dp_{1}}{dt}=\lambda p_{0}-\mu p_{1}\end{equation}
\\
However, since $p_{0}+p_{1}=1$, the redundancy in description
can be eliminated and by substituting $p_{1}=1-p_{0}$ in eq$^{n}$-2,
we describe \textbf{P} with respect to $p_{0}$ as :\\
\begin{equation}
\frac{dp_{0}}{dt}=\mu-(\lambda+\mu)p_{0}.\end{equation}
\\
Solving eq$^{n}$-4 for the initial condition $p_{0}(0)=1$, we
obtain :\\
\begin{equation}
p_{0}(t)=\frac{\mu}{\lambda+\mu}\left[1+\frac{\lambda}{\mu}e^{-(\lambda+\mu)t}\right]\end{equation}
\\
and therefore, \\
\begin{equation}
p_{1}(t)=\frac{\lambda}{\lambda+\mu}\left[1-e^{-(\lambda+\mu)t}\right]\end{equation}
\\
\\
To solve the next part of our query that is to find the probability
$p^{*}(t)$ that during any arbitrarily chosen time interval $(0,t)$
at least one part of \textbf{P} malfunctions at least once, we describe
\textbf{P} with a new set of states; viz. $(s_{0})$: when \textbf{P}
never fails, and $(s_{1})$: when at least one part of \textbf{P}
malfunctions at least once.\\
Here, solving the Chapman-Kolmogorov equation $\frac{dp_{0}}{dt}=-\lambda p_{0}^{*}$
for the initial condition $p_{0}^{*}(0)=1$, we get $p_{0}^{*}(t)=e^{-\lambda t}$
we arrive at the probability that during the time interval $(0,t)$
\textbf{P} malfunctions at least once; which is given by : $p_{1}^{*}(t)=1-p_{0}^{*}(t)=1-e^{-\lambda t}$.
\\
\\
To find the limiting probabilities, we study eq$^{n}$-5 and eq$^{n}$-6
when $t\rightarrow\infty$; whereby we arrive at the limiting probabilities
of the states, given by : $p_{0}=\frac{\mu}{\lambda+\mu}$ and $p_{1}=\frac{\lambda}{\lambda+\mu}$.\\
\\
\\
\textbf{\underbar{Case -2):}} \\
The idealistic framework described in case-1 may not always hold
true because, the malfunction in any part of \textbf{P} may not be
detected immediately but may take an interval of time. This non-trivial
interval of time is assumed to be represented by an exponential distribution
with some parameter $\theta$. Solving the Chapman-Kolmogorov equations
for the probabilities of the states for this case (along with the
limiting probabilities), will therefore be describing the biological
reality more realistically.\\
\\
For case-2 analysis, \textbf{P} is described with a set of three-states;
viz. $(s_{0})$: when \textbf{P} never fails and operates properly,
$(s_{1})$: when at least one part of \textbf{P} malfunctions at least
once but the malfunction is not detected, and $(s_{2})$: when at
least one part of \textbf{P} is being repaired.\\
\\
Denoting the corresponding probabilities for $(s_{0})$, $(s_{1})$
and $(s_{2})$ by $(p_{0})$, $(p_{1})$ and $(p_{2})$; the Chapman-Kolmogorov
equations for the probabilities of states can be constructed as :\\
\begin{equation}
\frac{dp_{0}}{dt}=\mu p_{2}-\lambda p_{0}\end{equation}
\\
\begin{equation}
\frac{dp_{1}}{dt}=\lambda p_{0}-\theta p_{1}\end{equation}
\\
and\\
\begin{equation}
\frac{dp_{2}}{dt}=\theta p_{1}-\mu p_{2}\end{equation}
\\
We convert the system of differential equations{[}7,8,9{]} to
an algebraic one by using Laplace transform. With due regard to the
intial conditions for transforms, say $\pi_{i}$ for probabilities
$p_{i}$, {[}7,8,9{]} the transformed system of equations can be represented
as {[}10,11,12{]}: \\
\begin{equation}
s\pi_{0}=\mu\pi_{2}-\lambda\pi_{0}+1\end{equation}
\\
\begin{equation}
s\pi_{1}=\lambda\pi_{0}-\theta\pi_{1}\end{equation}
\\
and\\
\begin{equation}
s\pi_{2}=\theta\pi_{1}-\mu\pi_{2}\end{equation}
\\
\\
Solving {[}10,11,12{]} algebraically we obtain :\\
\\
$\pi_{1}=\frac{\lambda}{s+\theta}\pi_{0}$ , $\pi_{2}=\frac{\theta}{s+\mu}\pi_{1}=\frac{\theta\lambda}{(s+\theta)(s+\mu)}\pi_{0}$
and finally, $\pi_{0}=\frac{(s+\theta)(s+\mu)}{s(s^{2}+s(\theta+\mu+\lambda)+\theta\lambda+\theta\mu+\lambda\mu)}$\\
\\
We denote :\\
\\
$a=\frac{\theta+\mu+\lambda}{2}+\sqrt{\frac{(\theta+\mu+\lambda)^{2}}{4}-\theta\lambda-\theta\mu-\lambda\mu}$\\
and\\
$b=-\frac{\theta+\mu+\lambda}{2}-\sqrt{\frac{\left(\theta+\mu+\lambda\right)^{2}}{4}-\theta\lambda-\theta\mu-\lambda\mu}$\\
\\
Whwreby the calculated probabilities can be expressed in closed
form as :\\
\begin{equation}
p_{0}(t)=\frac{ae^{at}-be^{bt}}{a-b}+\left(\theta+\mu\right)\frac{e^{at}-e^{bt}}{a-b}+\mu\theta\left[\frac{1}{ab}+\frac{be^{at}-ae^{bt}}{ab\left(a-b\right)}\right]\end{equation}
\\
\begin{equation}
p_{1}(t)=\lambda\frac{e^{at}-e^{bt}}{a-b}+\lambda\mu\left[\frac{1}{ab}+\frac{be^{at}-ae^{bt}}{ab\left(a-b\right)}\right]\end{equation}
\\
and\\
\begin{equation}
p_{2}\left(t\right)=\theta\lambda\left[\frac{1}{ab}+\frac{be^{at}-ae^{bt}}{ab\left(a-b\right)}\right]\end{equation}
\\
Advantage of such a scheme is that to evaluate the limiting probabilities
we can resort to either the transforms or the probabilities themselves
:\\
\\
\begin{equation}
p_{0}=\underset{t\rightarrow\infty}{lim}p_{0}\left(t\right)=\underset{s\rightarrow0}{lim}s\pi_{0}\left(s\right)=\frac{\mu\theta}{\lambda\mu+\lambda\theta+\theta\mu}\end{equation}
\\
\\
\begin{equation}
p_{1}=\frac{\lambda\mu}{\lambda\mu+\lambda\theta+\theta\mu}\end{equation}
 \\
and finally\\
\begin{equation}
p_{2}=1-p_{0}-p_{1}=\frac{\lambda\theta}{\lambda\mu+\lambda\theta+\theta\mu}.\end{equation}
\\
\\
\\
\textbf{References :}\\
Ahn, J., Kopelman, R., Argyrakis, P., 1999. Hierarchies of nonclassical
reaction kinetics due to anisotropic confinements. J. Chem. Phys.
110, 2116-2121.\\
Calef, D.F., Deutch, J.M., 1983. Diffusion-controlled reactions.
Ann. Rev. Phys. Chem. 34, 493-524.\\
Cox, B.G., 1994. Modern Liquid Phase Kinetics. Oxford University
Press, Oxford.\\
Ellis, R.J., 2001. Macromolecular crowding: obvious but underappreciated.
Trends Biochem. Sci. 26, 597-604.\\
Elowitz, M.B., Levine, A.J., Siggia, E.D., Swain, P.S., 2002.
Stochastic gene expression in a single cell. Science 297, 1183-1186.\\
Espenson, J.H., 1995. Chemical Kinetics and Reaction Mechanisms.
McGraw-Hill, Singapore.\\
Gibson M.A., 2000, Computational methods for stochastic biological
systems, PhD Thesis, California Institute of Technology, Pasadena,
California, 2000.\\
Gillespie D.T., 1977, Exact stochastic simulation of coupled chemical
reactions. J. Phys. Chem, 81:2340-2361.\\
Golightly A, Wilkinson DJ., 2006, Bayesian sequential inference
for stochastic kinetic biochemical network models., J Comput Biol.
13(3):838-851.\\
Hume, D.A., 2000. Probability in transcriptional regulation and
its implications for leukocyte differentiation and inducible gene
expression. Blood 96, 2323-2328. \\
Kuthan, H., 2001. Self-organisation and orderly processes by individual
protein complexes in the bacterial cell. Prog. Biophys. Mol. Biol.
75, 1-17.\\
Maamar H, Raj A and Dubnau D 2007 Noise in gene expression determines
cell fate in bacilus subtilis Science 317:526-529 \\
McAdams, HH, Arkin, A., 1999, Its a noisy business: Genetic regulation
at the nanomolar scale, Trends in Genetics 15,65-69.\\
Minton, A.P., 1993. Molecular crowding and molecular recognition.
J. Mol. Recognit. 6, 211-214.\\
Qian, H., Elson, E.L., 2002. Single-molecule enzymology: stochastic
Michaelis-Menten kinetics. Biophys. Chem. 101-102,565-576.\\
Rao C, Arkin A, 2003, Stochastic chemical kinetics and the quasi-steady-state
assumption: Application to the Gillespie algorithm. J Chem Phys 118:4999.
\\
Raser J. M., O'Shea E. K., 2004 Control of stochasticity in eukaryotic
gene expression Science 304:1811-1814.\\
Resat, H., Wiley, H.S., Dixon, D.A., 2001, Probability-weighted
dynamic Monte Carlo method for reaction kinetics simulations. J. Phys.
Chem. B 105, 11026-11034.\\
Sigman K., Discrete-time Markov chains, hypertext link: http://www.columbia.edu/\textasciitilde{}ks20/stochastic-I/stochastic-I-MCI.pdf\\
Turner T.E. , Schnell S., Burrage K., 2004, Stochastic approaches
for modelling in vivo reactions; Comp. Biol. and Chem. 28:165-178.\\
Weisstein, Eric W. \textquotedbl{}Chapman-Kolmogorov Equation.\textquotedbl{}
From MathWorld--A Wolfram Web Resource.\\
\\

\end{document}